# On the need for optimization of the software development processes in short-term projects

Moisés Homero Sánchez López[1], Carlos Alberto Fernández-y-Fernández[1] and Jorge Rafael Aguilar Cisneros [2]

[1] Instituto de Computación, Universidad Tecnológica de la Mixteca, carretera a Acatlima Km. 2.5 Huajuapan de León, Oax., México C.P. 69000,
`{moises, caff}@mixteco.utm.mx`

[2] Departamento de Ingenierías, Universidad Popular Autónoma del Estado de Puebla
21 sur #1103 Col. Santiago, Puebla Pue., México C.P. 72310
`jorge.aguilar@upaep.mx`

**Abstract.** Nowadays, most of the software development projects in Mexico are short-term projects (micro and small projects); for this reason, in this paper we are presenting a research proposal with the goal of identifying the elements contributing to their success or failure. With this research, we are trying to identify and propose techniques and tools that would contribute in the successful outcome of these projects.

## 1 Introduction

The appropiate use of techniques is a difficult job for micro, small and medium-sized business (MIPYMES)[1] engaged in software production [9]. Still, a considerable amount of software produced worldwide is built by MIPYMES, who often operate with limited resources both financial and time related. The vast majority of software producers, who do not use software development process, are paying high production costs and maintenance of systems, and therefore are being displaced in the domestic market, since they are not in a level of competitiveness as companies that use methods of developing and implementing process models [8].

Some software development businesses have been trying in recent years to incorporate best working practices for ensuring their productivity. This has also enabled them to deliver the quality demanded by the adoption of activities, practices, specific roles and norms of behaviour. Agile methods are mechanisms for software development that meet the specification of production speed and product quality [5]. As in most modern methodologies, agile methods are based in iterative and incremental development, encouraging deliveries in an evolutionary way and including modern values and practices in addition to the traditional ones. Authors and supporters of "the agile manifesto" ensure that, following its principles, it is possible to reduce risks and maintain precision [2]. In the agile methods, the planning takes place in each iteration and the plan is aligned with the business goals. In addition, the involvement of the client in the development ensures that the requirements are identified. By correctly using the "agile" philosophy it is possible to reach improvement goals of projects, efficiency and efficacy in the product, without affecting the quality. The agile methodology considers humane, organizational and technological aspects of the software development process [7].

---

[1] From its acronym in Spanish: Pequeñas y Medianas Empresas

One of these mechanisms for improvement is the systematic training, which ensures constant learning and application of new development tendencies. The key to success or failure of a software project depends strongly on solving the right problem [11].

On the other hand, there exist robust project management methodologies (not agile) such as the following:
- PMBOK, proposed by the Project Management Institute (PMI) [4] of the United States of America.
- PRINCE2, the project management method sponsored by the United Kingdom of Office of Government Commerce (OGC) [3].
- The International Project Management Association (IPMA) [13], represented in the United Kingdom for the qualification of the Association for Project Management (APM).

However, we do not believe that these methodologies contribute as expected to the success of the projects developed by micro and PyMEs. One of the researchers of this project was a full-time partner/director of a micro-company dedicated to software development for eleven years, and found that the problem raised in this paper is very common. Since 2006, this author has held informal talks with over thirty software development businesses from both CANACINTRA, Puebla and Huajuapan de León. In this period, the subject of short-term projects has been brought up repeatedly, where clients with a limited and fixed budget wish to automate some vital tasks from their businesses, however, They need the software immediately to get results as soon as possible.

The problem is that short-term projects, which are extremely constrained in cost and time, have a high probability of failure. In addition to this, there is no specific development methodology that takes into consideration the variables that determine the success of failure of them.

## 2 Project Overview

### 2.1 Problem Description

In the business environment of CANACINTRA, Puebla, companies offer customized development services to their clients on the condition of two major restrictions:
- Projects with budgets less than $50.000 (fifty thousand pesos).

- Projects with a timeline of less than twelve weeks for the delivery of full and put into operation project.

Short-term projects, with the features described above, are sometimes the only way for companies to justify the steady wage of its staff of programmers, and therefore develop these projects but with poor results or even catastrophic for both customer and provider.

It is then that the following questions are raised:
- Is it possible to deliver short-term projects with success, i.e. guarantee customer satisfaction as well as with revenues for the supplier?
- What are the characteristics that these projects must have?
- How can you measure the size of these projects?
- What are the requirements of the work team?
- What tools must developers use for each kind of project?
- What techniques and methodologies are appropriate for administration, requirements, design, development, and maintenance?
- What is the appropriate size for a work team?

The problem lies in analyzing a large sample of short-term projects to respond to the questions raised in the previous paragraph. We are seeking conclusions on the methods, strategies, techniques and tools that must be implemented according to the characteristics of short-term projects to ensure its success.

## 2.2 Justification

Between 80 and 90% of Mexican companies that develop software are micro and small companies MIPYMES (fewer than 50 employees) [6], of which, it is necessary to determine how many there are and how they often have been faced with the realization of short-term projects (this research is in progress). It might be thought that the short-term projects are responsible (hired) in turn by MIPYMES, however, in the experience of the authors, large organizations (more than 500 employees) are those who through some department with a specific need, require the implementation of short-term projects.

For this reason, it is necessary to identify the variables that maximize the probability of success, to ensure that the beneficiaries obtain the solution to their problems and needs with quality products, in a timely manner and within budget.

This study not only enhances the efficiency and competitiveness of small developers, but also will offer direct benefits to the contracting of the software projects. There are agile development methodologies such as extreme Programming (XP) [1], SCRUM [10, 12], test-driven development, among others, that offer short deliveries; however, those are not methodologies for such small projects, but rather, are aimed at making rapid and partial deliveries in projects of medium length (or long duration). However, it is very likely that under certain conditions, these same methodologies can be used to develop existing short-term projects.

## 2.3 Goals

The objective of this research is to identify and classify the factors that directly impact the success of short–term software projects, classifying projects based on the technology used, operating environment, as well as in the scope and size, to subsequently determine the methodologies, techniques and tools to develop this kind of projects, that will allow creating systems of limited size in a few weeks and with small teams of development. In addition, to identify the preconditions required according to the characteristics of the project and the number of weeks available for their development.

The specific objectives are:
- To determine the approximate number of short-term projects that small companies develop per year
- To obtain a sample of at least twenty short-term software projects from at least five different companies
- To analyze the factors that took the management process of the sample, from conception, recruitment, project management, product development, to the delivery
- To identify the characteristics that a short-term software project must have in order to be classified in this category and to be delivered successfully
- To find out what type of knowledge and skills successful development teams have had
- To know the tools that were used in the projects
- To develop a ranking of projects according to its technological features and functionalities, identifying the variables that influenced each type
- To determine the variables that have been a factor of success or failure in each of the sample projects
- To determine the degree of incidence for each variable
- To develop a hypothesis about the degree of importance and influence of each variable in each of the types of projects, testing the hypothesis in projects that are in progress in the last phase of this research
- To determine the tools that must be made available to the development team
- To propose a set of techniques and/or process model for the development of short-term software projects

## 3 Methodology

Our research is divided into three stages. The first stage will be basically for the study of our target population. The second stage will be training and analysis. In the last stage we'll be proposing a set of techniques and tools based on what has been achieved in the previous stages. Figure 1 shows in general terms the proposed methodology described in this section. Below the stages of the project are explained in detail.

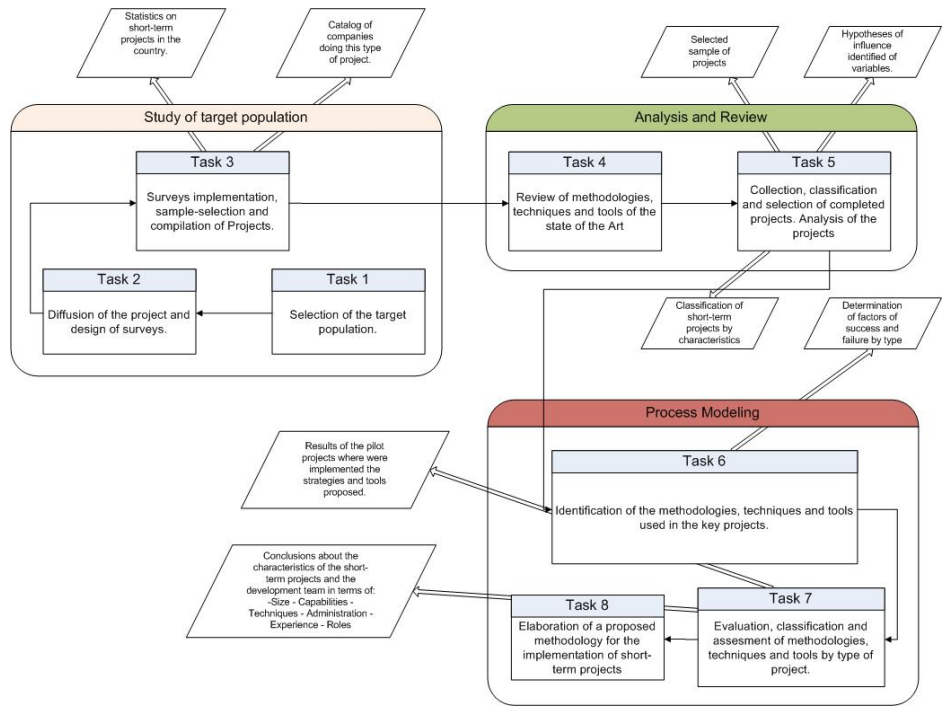

**Figure 1. Methodology of the research**

## 3.1 Stage 1: Study of the target population

At this stage, an announcement about the research project will be made to a group of over one hundred companies inviting them to participate by granting of their projects to the research team. Surveys will be carried out by executives, in order for the team to know the approximate number of short-term projects that are carried out in an annual period.

At this stage a representative figure will be required on the number of short-term projects that are carried out in the country for measuring the benefit this research will have. In addition, companies that show a real interest in this project will provide a sample of short-term projects that have already been completed. It is important to know the impact that this investigation may have if its results are disseminated in the business community of Information Technologies.

At the end of this stage we will have reliable statistics on the number of short-term projects that are carried out in the country. In addition, we will have a catalog of companies doing these types of projects and their level of interest in the results of the investigation.

## 3.2 Stage 2: Training and Analysis

In this stage the research team will be trained to manage projects and to develop software using different methodologies and techniques. Training will be considered according to the results of the previous stage. In this stage we will collect companies' projects in order to carry out their pre-analysis. After that, we will analyze the projects to determine variables that influence their success or failure. The analysis will take into account:
- project size (modules), functionality and scope
- system technological infrastructure
- technology used for development
- project services enclosed
- management methodology used
- development methodology used
- tools used for management and development
- cost and time performance project analysis
- technical capabilities and team experience
- team size
- team roles assigned

When this stage is complete, we will choose at least twenty software projects and generate a list indentifying the most important characteristics of the projects. We will identify the key factors that directly influence project performance. We will create a comparative framework about methodologies and tools with the purpose of identifying the best option.

In order to select the projects that will be the subject of the research, we will have to analyze and study projects provided by the participating companies, the selection criteria will be:

- size and characteristics of the project, i.e. it will be validated that is a complete project, which has required analysis, design, construction, testing and delivery, and which has been obtained through a formal contract as well as has required administration;
- delivery time of projects, the research team considers short-term projects to those with delivery times under or around 12 weeks;
- the size of the team: having less than three programmers is ideal to be considered short-term project.

## 3.3 Stage 3: Process Model

At stage three, we will select the best methodologies, strategies, techniques and tools for short-term project development. We will do knowledge transfer to test companies' development/management teams. We will test our proposals on some projects which are starting. Finally, we will model the development of activities from a process-based approach.

The goal is to generate a proposal list to develop short-term projects:
- administration methodologies
- development methodologies
- techniques
- tools
- process models improvement

Our proposals may be used as a guide to increase the competitive advantage of participating companies. Micro and small companies will ensure the return on investment of their money. Large companies could control their software development suppliers. The principal beneficiary will be the customer.

## 4 Partial outcomes

A national survey was disseminated and conducted with the support of the *Secretaría de Economía and Insitituto Mexicano de Desarrollo de Software*. There was a total of 107 participating companies. The survey data are being analyzed and will be presented in future articles. At the same time, we are conducting an analysis of software projects of some companies to determine their degree of complexity. To do this, we are using function points analysis. By analyzing projects, we can identify key factors which contribute to its success or failure.

**Acknowledgments**. This work has been funded by the Universidad Tecnológica de la Mixteca (UTM). The survey was conducted by the UPAEP (Universidad Popular Autónoma del Estado de Puebla). Stage one of the project also has the economical support at PROMEP.